\newcommand\aj{{AJ\,}}%
\newcommand\apj{{ApJ\,}}%
\newcommand\apjl{{ApJ\,}}%
\newcommand\apjs{{ApJS\,}}%
\newcommand\aap{{A\&A\,}}%
\newcommand\mnras{{MNRAS\,}}%
\newcommand\nat{{Nature\,}}%

\documentclass[graybox, natbib, footinfo]{svmult}

\usepackage{url}

\usepackage{mathptmx}       
\usepackage{helvet}         
\usepackage{courier}        
\usepackage{type1cm}        
\usepackage{makeidx}         
\usepackage{graphicx}        
\usepackage{multicol}        
\usepackage[bottom]{footmisc}

\usepackage{journals}

\makeindex             

\begin{document}
\title*{Effects of rotation and thermohaline mixing in red giant stars}
\author{Corinne Charbonnel, Nad\`ege Lagarde  \& Patrick Eggenberger}
\institute{C. Charbonnel \at Observatoire de Gen\`eve, Universit\'e de Gen\`eve, 51 ch. des Maillettes 1290 Versoix (Switzerland), \\ Laboratoire d'Astrophysique de Toulouse-Tarbes, CNRS UMR 5572, Universit\'e de Toulouse, 14 av. E.Belin F-31400 Toulouse, \\ \email{Chabonnel.Corinne@unige.ch}
\and N. Lagarde \at Observatoire de Gen\`eve, Universit\'e de Gen\`eve, 51 ch. des Maillettes 1290 Versoix (Switzerland), \\ \email{Lagarde.Nadege@unige.ch}
\and P. Eggenberger. \at Observatoire de Gen\`eve, Universit\'e de Gen\`eve, 51 ch. des Maillettes 1290 Versoix (Switzerland), \\ \email{Patrick.Eggenberger@unige.ch}}
%
%
\maketitle

\abstract{Thermohaline mixing has been recently identified as the probable dominating process that governs the photospheric composition of low-mass bright giant stars \citep{ChaZah07a}.
Here we present 
the predictions of stellar models computed with the code STAREVOL including this mechanism together with rotational mixing. We compare our theoretical predictions with recent observations.}

\section{Introduction}
\label{sec:1}

The standard theory of stellar evolution predicts that the convective envelope of low-mass stars deepens in mass during the contraction of the degenerate He-core after the main sequence turnoff, and engulfes hydrogen-processed material (the so-called first dredge-up, hereafter 1dup). This induces a decrease of the surface $^{12}$C/$^{13}$C ratio and of the Li and $^{12}$C abundances, while $^{14}$N and $^{3}$He abundances increase. After the 1dup, the convective envelope retracts and the hydrogen burning shell (HBS) moves outward (in mass). According to the standard theory,  no further change of the surface chemical properties is expected after the 1dup on the Red Giant Branch (RGB). 
However, spectroscopic observations \citep{Gilroy89,GiBr91,Luck94,Gratton00,Tautvaisiene00,Tautvaisiene05,Smith02,Shetrone03,Pilachowski03,Geisler05,Spite06,ReLa07,Smiljanic09} show clear signatures of ``extra-mixing" on the upper RGB in low-mass stars, when the HBS crosses the discontinuity left behind by the 1dup at the bump luminosity. More specifically, the carbon isotopic ratio and the C and Li abundances drop again as the stars move across the bump. Different processes have been proposed to explain these abundance anomalies.
Here we  recall the potential impact of rotation-induced mixing and thermohaline instability on the RGB.  

\section{Rotation-induced mixing}

\begin{figure}
\vspace*{+1.5cm}
\hspace{0.3cm}
\includegraphics[angle=0,width=7cm,trim= 2.5cm 1cm 1cm 2cm]{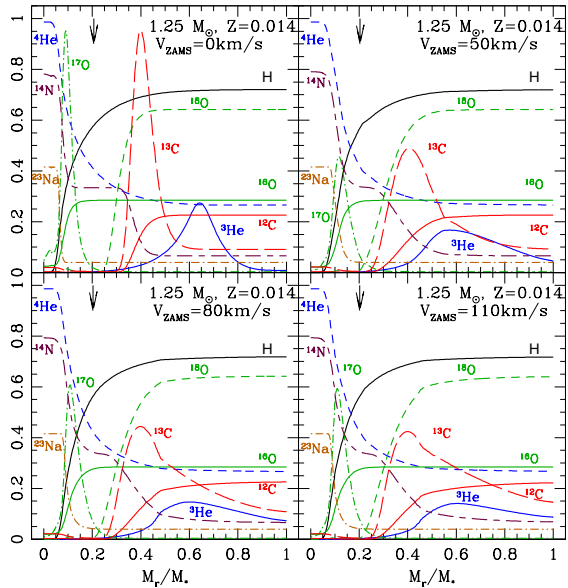}\hfill
\begin{minipage}[r]{0.33\linewidth}
\vspace*{-6.5cm}
\caption{Chemical structure at the turnoff of a 1.25~M$_{\odot}$ star computed for different initial rotation velocities as indicated. The mass fractions are multiplied by 100 for $^3$He, $^{12}$C, and $^{14}$N, by 2500 for $^{13}$C, by 50, 900, and 5$\times 10^4$ for $^{16}$O, $^{17}$O, and $^{18}$O respectively, and by 1500 for $^{23}$Na.
	  The vertical arrows show, in all cases, the maximum depth reached by the convective envelope at its maximum extent during the 1dup. Figure from \cite{ChLa10}.}
\label{fig:1}       
\end{minipage}\hfill
\end{figure}

Rotation-induced mixing has an impact on the internal abundance profiles of chemicals involved in hydrogen-burning while the stars are on the main sequence \cite[][]{TalCha98, TaCC10, ChaTal99,Palacios03,Pasquini04,Smiljanic09,ChLa10}.
This can be seen in Fig. \ref{fig:1} where we present the chemical structure at the end of the main sequence for $1.25~M_{\odot}$ stellar models computed assuming different initial rotation velocities. In the rotating models, the abundance gradients are smoothed out compared to the standard (i.e., non-rotating) case:  $^{3}$He, $^{13}$C, $^{14}$N,
and $^{17}$O diffuse outwards, while $^{12}$C and $^{18}$O diffuse inwards. After the turnoff , this reflects in different post-dredge up predictions for the surface abundances that agree well with observations in low-luminosity red giant stars (see Fig.2).  However, rotation-induced mixing does not explain the abundance anomalies observed in low-mass red giants brighter than the bump \citep[][]{Chaname05,Palacios06}. 

\section{Thermohaline mixing}
\label{sec:2}

\cite{ChaZah07a} identified thermohaline mixing as the process that governs the photospheric composition of low-mass stars above the bump on the RGB. At this evolutionary phase, this double diffusive instability is induced by the $^{3}$He($^{3}$He; 2p)$^{4}$He reaction \citep{Eggleton06} that creates an inversion of mean molecular weight  \citep{Ulrich71}.
Here we use the following prescription for the turbulent diffusivity coefficient \citep{Ulrich72,Kippen80} : 

\begin{equation}
{\rm D_ t} =  {\rm C_ t} \,  {\rm K}  \left({\varphi \over \delta}\right){- \nabla_\mu \over (\nabla_{\rm ad} - \nabla)} \quad \hbox{for} \;  \nabla_\mu < 0, 
\label{dt}
\end{equation}
 
\begin{equation}
{\rm C_ t} = {8 \over 3} \pi^2 \alpha^2 ,  
\end{equation}
with K the thermal diffusivity and $\alpha = 5$ the aspect ratio of salt fingers \citep{Ulrich72}.
At the RGB bump and above, the thermohaline diffusion coefficient  is several order of magnitudes higher than the total diffusion coefficient related to rotation-induced processes \citep{ChLa10}.

\section{Model predictions and comparisons to observations.}
\label{sec:3}

The results shown in this section are presented and discussed in detail in \cite{ChLa10} and have been computed with the code STAREVOL including rotation-induced processes and thermohaline mixing. Here we briefly discuss the effects of rotation-induced mixing and thermohaline mixing on carbon isotopic ratio, lithium and $^{3}$He.

\subsection{Carbon isotopic ratio}

\begin{figure}
\includegraphics[angle=0,width=7.5cm,trim = 1cm 1cm 1cm 1cm]{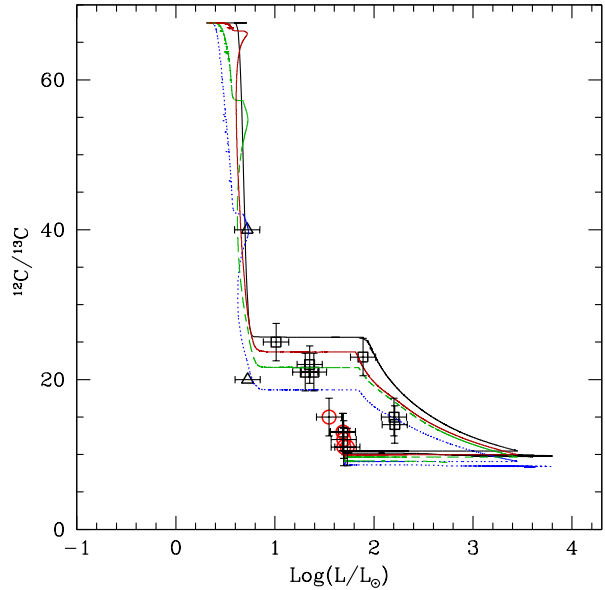}\hfill
\begin{minipage}[r]{0.33\linewidth}
\vspace*{-6.5cm}
\caption{Evolution of the surface $^{12}$C/$^{13}$C ratio as a function of stellar luminosity for models of  1.25~M$_{\odot}$ stars. 
Different tracks are for different initial rotation velocities (non-rotating case, 50, 80, and 110~km.s$^{-1}$ respectively in black, red, green, and blue). Observations  by \cite{GiBr91} in evolved stars of the open cluster M67 (turnoff mass $\sim$1.2~M$_{\odot}$) are also shown (triangle, squares, and circles for subgiant, RGB, and clump stars respectively). Figure from \cite{ChLa10}.}
\label{fig:2}       
\end{minipage}\hfill
\end{figure}

In Fig.\ref{fig:2} we compare the evolution of the theoretical  surface carbon isotopic ratio for $1.25~M_{\odot}$ models  with observations in M67 stars by \cite{GiBr91}. We note that rotation-induced mixing on the main sequence slightly lowers the post-dredge-up $^{12}$C/$^{13}$C value (see \S2).
On the other hand, thermohaline mixing leads to further decrease of the carbon isotopic ratio at the luminosity of the bump ($\log(L/L_{\odot})\sim$2), in excellent agreement with M67 data.

\begin{figure}
\vspace*{0.5cm}
\includegraphics[angle=0,width=7.5cm,trim= 0.5cm 0.5cm 2cm 2cm ]{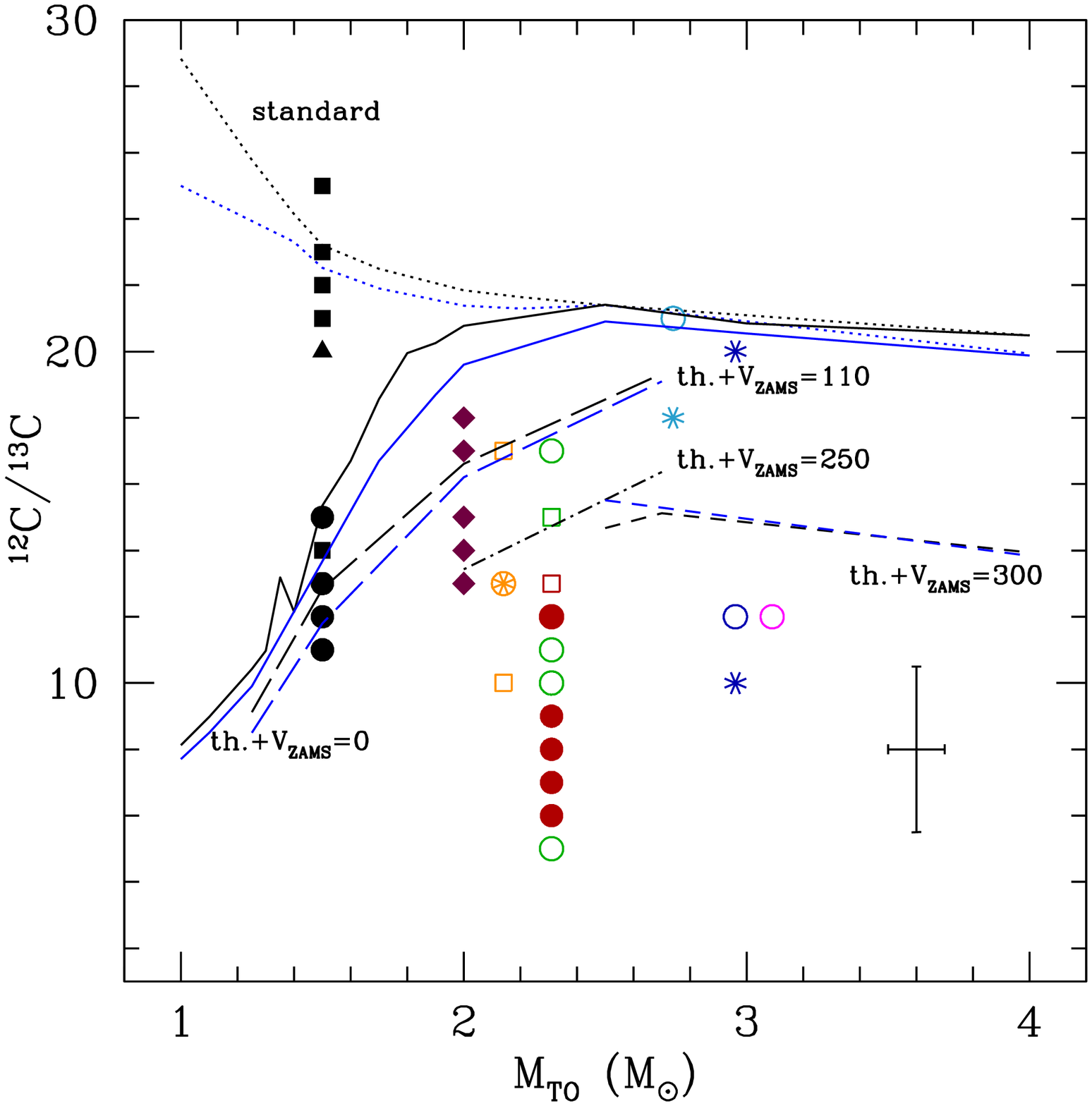}\hfill
\vspace*{.5cm}
\caption{Observations of $^{12}$C/$^{13}$C in evolved stars of Galactic open clusters by  \cite{Smiljanic09},  \cite{Gilroy89}, \cite{GiBr91}, and \cite{Mikolaitisetal10} as a function of the turnoff mass of the corresponding host cluster.
	  Squares, circles, and asteriscs are for RGB, clump, and early-AGB stars respectively, while diamonds are for stars from \cite{Gilroy89} sample with doubtful evolutionary status; triangles are for lower limits. 
	  	  A typical error bar is indicated. 
	  Theoretical predictions are shown at the tip of the RGB and after completion of the second dredge-up (black and blue lines respectively). Standard models (no thermohaline nor rotation-induced mixing) are shown as dotted lines, models with thermohaline mixing only (V$_{\rm ZAMS}$=0) as solid lines, and models with thermohaline and rotation-induced mixing for different initial rotation velocities as indicated as long-dashed, dot-dashed, and dashed lines.
Figure from \cite{ChLa10}. }
\label{fig:3}       
\end{figure}

In Fig.\ref{fig:3} we show the predictions for the surface $^{12}$C/$^{13}$C ratio at the tip of the RGB and at the end of second dredge-up (black and blue lines respectively) for models over a large mass range and compare them with observations in stars belonging to open clusters of various turnoff masses. We see that in stars with masses below $\sim$2M$_{\odot}$,
thermohaline mixing is the main physical process governing the photospheric composition of evolved giants, although the final carbon isotopic ratio also slightly depends on rotation-induced mixing on the main sequence.
In intermediate-mass stars that do not reach the bump on the RGB and do not undergo thermohaline mixing at that phase, rotation is necessary to explain the data and accounts for star-to-star abundance variations at a given evolutionary status.
Overall, the present models explain very well the observed abundance patterns over the considered mass range.

\subsection{Lithium}

\begin{figure}
\vspace*{.5cm}
\includegraphics[angle=0,width=7.5cm,trim= 1cm 1cm 1cm 1cm]{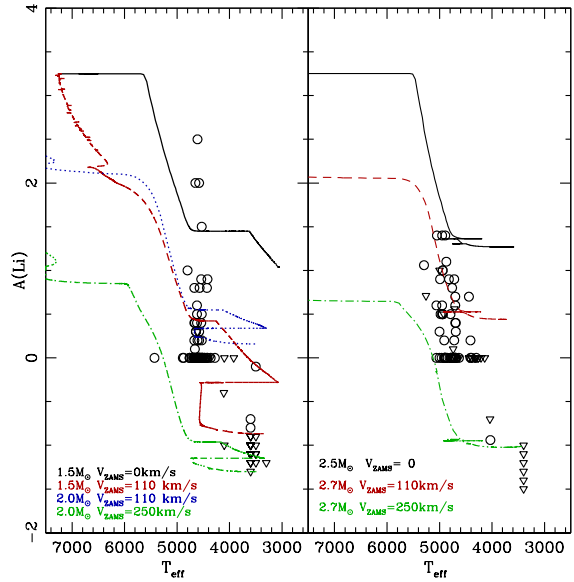}\hfill
%
%
\begin{minipage}[r]{0.33\linewidth}
\vspace*{-7.5cm}
\caption{Lithium data for field evolved stars from the sample by Charbonnel et al. (in prep.) that are segregated according to their mass (left and right panels include respectively sample stars with masses lower and higher than 2~M$_{\odot}$; Li detections and upper limits are shown as circles and triangles respectively). 
	  Theoretical lithium evolution is shown from the ZAMS up to the end of the early-AGB. 
	  Various lines correspond to predictions for stellar models of different masses computed without or with rotation as indicated, and with thermohaline mixing in all cases.}
\label{fig:4}       
\end{minipage}\hfill
\end{figure}

In Fig.\ref{fig:4} we present lithium data for field red giant stars with metallicities around solar and precise determination of their evolutionary status, and compare them to predictions for models of various masses.  
Contrary to the standard case, rotation-induced mixing leads to Li depletion on the main sequence \cite[see e.g.][]{TalCha98,TaCC10,Palacios03}.
After the 1dup, the surface Li abundance  remains constant  until the stars reach the bump luminosity where thermohaline mixing becomes efficient and destroys Li. Later on the star reaches the RGB tip, and finally the second dredge-up decreases again Li at the surface. Models are in perfect agreement with observations. 

\begin{figure}
\vspace*{0.5cm}
\includegraphics[angle=0,width=7.5cm,trim= 1cm 1cm 1cm 1cm]{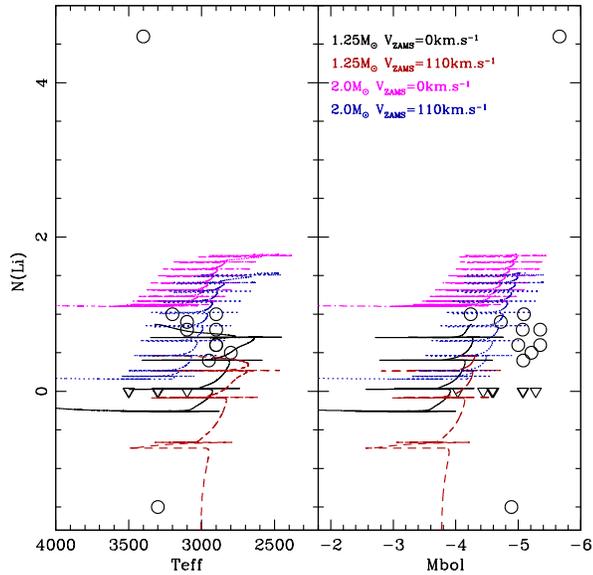}\hfill
%
%
\begin{minipage}[l]{0.33\linewidth}
\vspace*{-7.5cm}
\caption{Lithium observations in oxygen-rich variables belonging to the Galactic disk  as a function of effective temperature and bolometric magnitude \citep{UttLeb10}; circles and triangles are for abundance determinations and upper limits respectively. 
 Theoretical lithium evolution is shown from the early-AGB up to the end of the TP-AGB. 
 Various lines correspond to predictions for stellar models of different masses computed without or with rotation as indicated, and with thermohaline mixing in all cases. Figure from \cite{ChLa10}.}
\label{fig:5}       
\end{minipage}\hfill
\end{figure}

We have computed a few models along the TP-AGB, and found that thermohaline mixing leads to non negligible fresh lithium production, as shown in Fig.\ref{fig:5}.
There we present the evolution of the surface lithium abundance as a function of both effective temperature and bolometric magnitude for TP-AGB models of 1.25 and 2.0M$_{\odot}$ stars.
Theoretical predictions are compared with observations of the sample of low-mass oxygen-rich AGB variables belonging to the Galactic disk studied by \cite{UttLeb10}, and are found to fit very nicely the observed Li behaviour. Let us note that despite the strong production of fresh Li at that phase, the total stellar yields remain negative for this element. 

\subsection{Helium 3} 

\begin{figure}
\vspace*{1cm}
\hspace{0.3cm}
\includegraphics[angle=0,width=7.5cm,trim= 1cm 1cm 1cm 1cm]{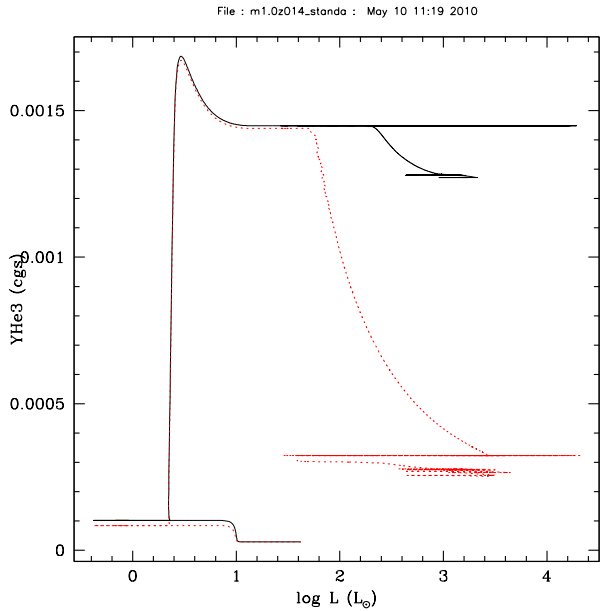}\hfill
\begin{minipage}[r]{0.28\linewidth}
\vspace*{-5cm}
%
%
\caption{Evolution of the surface abundance of $^{3}$He (in mass fraction) from the pre-main sequence up to the AGB tip for 1M$_{\odot}$ models at solar metallicity. The black solid line and the red dotted-line correspond to the standard and thermohaline cases respectively. }
\label{fig:6}       
\end{minipage}\hfill
\end{figure}

On the main sequence, a $^{3}$He peak builds up due to pp-reactions inside low-mass stars \citep{Iben67}, and is engulfed in the stellar envelope during the 1dup. As a consequence the surface abundance of $^{3}$He strongly increases on the lower RGB as can be seen in Fig.\ref{fig:6} which presents the evolution of $^{3}$He mass fraction at the surface of  $1~M_{\odot}$ model at solar metallicity in the standard case and in the case with thermohaline mixing (black solid and red dotted lines respectively). After the bump, thermohaline mixing transports $^{3}$He from the convective envelope down to the hydrogen-burning shell where it burns. This leads to a rapid decrease of the surface abundance (and thus of the corresponding yield) of this element as can be seen in Fig.\ref{fig:6}. We are presently computing similar stellar models over a large range in stellar mass and metallicity  in order to quantify the actual contribution of low-mass stars to Galactic $^{3}$He (Lagarde et al., in preparation). We are confident that the corresponding $^{3}$He yields will help reconciling the primordial nucleosynthesis with measurements of $^{3}$He/H in Galactic HII regions \citep{Charbonnel02}.

\section{Conclusions}

An inversion of molecular weight created by the $^{3}$He($^{3}$He; 2p)$^{4}$He reaction is at the origin of thermohaline instability in low-mass RGB stars brighter than the bump. The associated mixing explains very well observations of $^{12}$C/$^{13}$C and Li in these objects. Rotation-induced mixing, coupled with thermohaline mixing, allows us to explain the $^{12}$C/$^{13}$C and Li data in giant stars over a large mass range. Thermohaline mixing has also an effect during the TP-AGB phase, where it leads to fresh lithium 
production.  Finally, this process helps reconciling stellar yields predictions with the Galactic evolution of $^3$He as constrained by the data in Galactic HII regions.

\begin{acknowledgement}
We acknowledge financial support from the Swiss National Science Foundation (FNS) and from the french Programme National Program (PNPS) of CNRS/INSU.
\end{acknowledgement}

\end{document}